\documentclass[fleqn,usenatbib]{mnras}
\usepackage[T1]{fontenc}
\usepackage{ae,aecompl}

\usepackage{graphicx}	
\usepackage{amsmath}	
\usepackage{amssymb}	

\usepackage{ulem}

\usepackage{xcolor}
\usepackage{ragged2e}

\title[Model for IC 342 X-1 source]
	{Relativistic viscous accretion flow model for ULX sources: A case study for IC 342 X-1 }

\author[Das et al.]{Santabrata Das$^{1}$\thanks{E-mail: sbdas@iitg.ac.in (SD)}, 
Anuj Nandi$^{2}$\thanks{E-mail: anuj@ursc.gov.in (AN)}, Vivek K. Agrawal$^{2}$, Indu Kalpa Dihingia$^{3}$,
\newauthor Seshadri Majumder$^{1}$\\
$^1$ Indian Institute of Technology Guwahati, Guwahati, 781039, Assam, India\\
$^2$ Space Astronomy Group, ISITE Campus, U. R. Rao Satellite Center, Outer Ring Road, Marathahalli, Bangalore 560037, India\\
$^3$ Discipline of Astronomy, Astrophysics and Space Engineering, Indian Institute of Technology Indore, Indore 453552, India
}

\date{Accepted XXX. Received YYY; in original form ZZZ}

\begin{document}
\label{firstpage}
\pagerange{\pageref{firstpage}--\pageref{lastpage}}
\maketitle

\begin{abstract}
		In this letter, we develop a model formalism to study the structure of a relativistic, viscous, optically thin, advective accretion flow around a rotating black hole in presence of radiative coolings. We use this model to examine the physical parameters of the Ultra-luminous X-ray sources (ULXs), namely mass ($M_{\rm BH}$), spin ($a_{\rm k}$) and accretion rate (${\dot m}$), respectively. While doing this, we adopt a recently developed effective potential to mimic the spacetime geometry around the rotating black holes. We solve the governing equations to obtain the shock induced global accretion solutions in terms of ${\dot m}$ and viscosity parameter ($\alpha$). Using shock properties, we compute the Quasi-periodic Oscillation (QPO) frequency ($\nu_{\rm QPO}$) of the post-shock matter (equivalently post-shock corona, hereafter PSC) pragmatically, when the shock front exhibits Quasi-periodic variations. We also calculate the luminosity of the entire disc for these shock solutions. Employing our results, we find that the present formalism is potentially promising to account the observed $\nu_{\rm QPO}$ and bolometric luminosity ($L_{\rm bol}$) of a well studied ULX source IC 342 X-1. Our findings further imply that the central source of IC 342 X-1 seems to be rapidly rotating and accretes matter at super-Eddington accretion rate provided IC 342 X-1 harbors a massive stellar mass black hole ($M_{\rm BH} < 100 M_\odot$) as indicated by the previous studies.
	
\end{abstract}



\section{Introduction}
\label{sec:intro}

Since discovery, ULXs draw significant attention among the researchers due to its exceedingly high luminosity $\sim 10^{39-40}$ erg s$^{-1}$ \citep{Fabbiano1989}. The true nature of the central accretor of ULXs and the exact physical mechanism responsible for such a high luminosity still remain elusive. Meanwhile, different competing ideas gain popularity to elucidate this. First possibility assumes the ULXs to harbor stellar mass black holes that accrete at super-Eddington rate \citep{Fabrik-Mescheryakov2001,Poutanen-etal2007}. Second possibility considers stellar mass black hole X-ray binaries (XRBs) accreting at sub-Eddington rate with beamed emission \citep{Reynolds-etal1997,King2002}, although the observational evidence of beaming effect is not well understood  \citep{Feng-Soria2011}. The third alternative scenario presumes the central source to be the intermediate mass black holes (IMBHs) of mass $10^{3} - 10^{5} M_\odot$ \citep{Colbert-Mushotzky1999,Makishima-etal2000} that accrete at sub-Eddington accretion rate while emitting high luminosity. Needless to mention that all these models are in contrast and therefore, remain inconclusive.

So far, numerous efforts were made to constrain the mass of the ULX sources through the spectral and timing studies
\citep{Watarai-etal2001,Dewangan-etal2006,Pasham-etal2015,Agrawal-Nandi2015,Kaaret-etal2017,Mondal-etal2020,Ghosh-Rana2021}. Furthermore, the presence of super-Eddington accretion rate for several ULXs is also reported \citep{Gladstone-etal2009} which implies a new accretion state named the ultraluminous state. In parallel, efforts were also given in the theoretical front, where models are developed to examine the observational signature of ULXs \citep[and references therein]{Middleton-etal2015,Mondal-Mukhopadhyay2019,Middleton-etal2019}. Indeed, the investigation of the physical parameters (mass and spin) of the central sources remain unexplored in these works.

Motivating with this, in this letter, we investigate $L_{\rm bol}$ and $\nu_{\rm QPO}$ of ULXs adopting a relativistic, viscous, advection dominated accretion flow model around the rotating black holes in presence of cooling. To validate our model formalism, we consider a ULX source IC 342 X-1 for the purpose of representation, and compute the possible ranges of $M_{\rm BH}$, $a_{\rm k}$ and ${\dot m}$ that yields the observed $\nu_{\rm QPO}$ and $L_{\rm bol}$, simultaneously.

The letter is organized as follows. In \S 2, we present the underlying assumptions and model equations that describe the flow motion. In \S 3, we discuss the accretion solutions and compute the observables. In \S 4, we present the observational features of IC 342 X-1 source and constrain the physical parameters of the source using our model formalism. Finally, we conclude with discussion in \S 5.

\section{Assumptions and Model Equations}
 
\label{sec:model}
 
We consider a relativistic, steady, viscous, optically thin, advective accretion disc around a ULX source. To describe the spacetime geometry around the central object, we adopt a newly formulated effective potential \citep{Dihingia-etal2018}. We express the flow variables in dimensionless unit by considering an unit system $G=M_{\rm BH}=c=1$, where $G$, $M_{\rm BH}$, and $c$ are the gravitational constant, black hole mass, and speed of light, respectively. In this unit system, radial distance is expressed in unit of $r_g=GM_{\rm BH}/c^2$. We use cylindrical coordinate system keeping the central source at the origin.
 
We develop a model of accretion flow where the governing equations that describe the flow structure are given by \citep{Chakrabarti1996},
\begin{eqnarray}
&&u\frac{du}{dr}+\frac{1}{h\rho}\frac{dP}{dr}  + \frac{\partial \Phi_{\rm eff}}{\partial r}
= 0,\\
&&u\frac{d\lambda}{dr} + \frac{1}{\Sigma x}\frac{d}{dr}\left(r^2 W_{r \phi}\right)=0,\\
&&\dot{M} = 2\pi u \Sigma \sqrt\Delta,\\
&&\Sigma u T\frac{ds}{dr} = \frac{Hu}{\Gamma-1}\left(\frac{dP}{dr}-\frac{\Gamma P}{\rho}\frac{d\rho}{dr} \right)=Q^{+}-Q^{-},
\end{eqnarray}
where $r$, $u$, $h$, $P$, and $\rho$ are the radial coordinate, radial velocity, specific enthalpy, isotropic pressure, and density, respectively. The effective potential is given by,
$
\Phi_{\rm eff}= \frac{1}{2} \ln\bigg[\frac{r\Delta}{a_{\rm k}^2 (r+2)-4 a_{\rm k} \lambda +r^3-\lambda ^2 (r-2)}\bigg]
$
 \citep{Dihingia-etal2018}, where $\lambda$ is the specific angular momentum of the flow, $a_{\rm k}$ is the Kerr parameter, and $\Delta=r^2 - 2r + a_{\rm k}^2$. In Eq. (2), the viscous stress $W_{r\phi}=\alpha (W+\Sigma u^2)$ \citep{Chakrabarti-Das2004}, where $\alpha$ refers viscosity parameter, $W$ is the vertically integrated pressure, $\Sigma ~(=\rho H)$ is the surface mass density, and the vertical disc height $H=\sqrt{P{\cal F}/\rho}$, $\mathcal{F}=\left(1-\Omega\lambda\right)r^3\left[(r^2 + a_{\rm k}^2)^2 - 2\Delta a_{\rm k}^2\right]\left[(r^2 + a_{\rm k}^2)^2 + 2\Delta a_{\rm k}^2\right]^{-1}$, with $\Omega$ being the angular velocity of the flow \citep{Riffert-Herold1995,Peitz-Appl1997}. In Eq. (3), $\dot{M}$ denotes accretion rate which is expressed in dimensionless form as ${\dot m}={\dot M}/{\dot M}_{\rm Edd}$, where $\dot{M}_{\rm Edd}=1.44\times10^{17}\left( \frac{M_{\rm BH}}{M_{\odot}}\right)$ g s$^{-1}$. In Eq. (4), $s$ is the specific entropy, $T$ is the temperature, $Q^{+}\ \left[=-\alpha r\left(W + \Sigma u^2\right)\frac{d\Omega}{dr}\right]$ \citep{Chakrabarti-Das2004} is the heating due to viscous dissipation, and $Q^{-} \ \left[=Q_{\rm b} + Q_{\rm cs} + Q_{\rm mc}\right]$ \citep{Mandal-Chakrabarti2005} is the energy loss through radiative coolings, where $Q_{\rm b}$, $Q_{\rm cs}$, and $Q_{\rm mc}$ are for bremsstrahlung, cyclo-synchrotron, and Comptonization processes. Following \citet{Chattopadhyay-Chakrabarti2000}, we compute electron temperature as $T_e=\sqrt{m_e/m_p}T$, where $m_e$ and $m_p$ are the masses of electron and ion. In this work, we employ equipartition to calculate magnetic fields ($B$) for simplicity and obtain as $B=\sqrt{8 \pi \beta P}$, where $\beta = 0.1$ is assumed \citep{Mandal-Chakrabarti2005}.

Governing equations (1-4) are closed with an equation of state (EoS), which we choose for relativistic flow as $e = n_em_ef=\rho f /\tau$ \citep{Chattopadhyay-Ryu2009}, where $f = (2-\xi)\left[1 + \Theta\left(\frac{9\Theta + 3}{3\Theta + 2}\right)\right] +  \xi \left[ \frac{1}{\chi} + \Theta\left(\frac{9\Theta + 3/\chi} {3\Theta + 2/\chi}\right)\right]$, $\tau=[2-\xi(1-1/\chi)]$, $\chi = m_e/m_p$, $\xi = n_p/n_e$, $\Theta=k_{\rm B}T/m_ec^2$, $k_{\rm B}$ is the Boltzmann constant, and $n_e~(n_p)$ denotes the number density of the electron (ion). With this, we express the polytropic index $N = \frac{1}{2}\frac{df}{d\Theta}$, the ratio of specific heats $\Gamma = 1 + \frac{1}{N}$ and the sound speed $a_s^2 = \Gamma p/\left(e+p\right) = 2\Gamma\Theta/\left( f + 2\Theta\right)$. 
Here, we assume $\xi=1$ unless stated otherwise.

We study the global accretion solutions around a ULX source following the methodology described in \cite{Chakrabarti-Das2004}, where the basic equations (1-4) are simultaneously integrated for a specified set of flow parameters. To do this, we treat $\alpha$, ${\dot m}$, and $M_{\rm BH}$ as global parameters. Because of the transonic nature of the equations, at the inner critical point $r_{\rm in}$, we choose the boundary values of angular momentum ($\lambda_{\rm in}$) and energy (equivalently Bernoulli parameter ${\cal E}_{\rm in}= \left[ u^2/2 + \log h + \Phi_{\rm eff}\right]_{r_{\rm in}}$) of the flow as local parameters \citep{Dihingia-etal2020}. Using these parameters, equations (1-4) are integrated starting from $r_{\rm in}$ once inward up to just outside the horizon and then outward up to a large distance (equivalently `outer edge of the disc') to get the complete accretion solution. Depending on the input parameters, accretion flow may possess multiple critical points \citep{Das-etal2001a} and also experience shock transitions provided the shock conditions are satisfied \citep{Landau-Lifshitz1959,Fukue-1987,Fukue-2019a,Fukue-2019}. We compute the shock induced global accretion solutions as these solutions are potentially viable to explain the observational findings of black hole X-ray sources \citep[and references therein]{Chakrabarti-Titarchuk1995,Iyer-etal2015,Sreehari-etal2019}.

\section{Accretion solutions and observables}

\begin{figure}
	\begin{center}
		\includegraphics[scale=0.4]{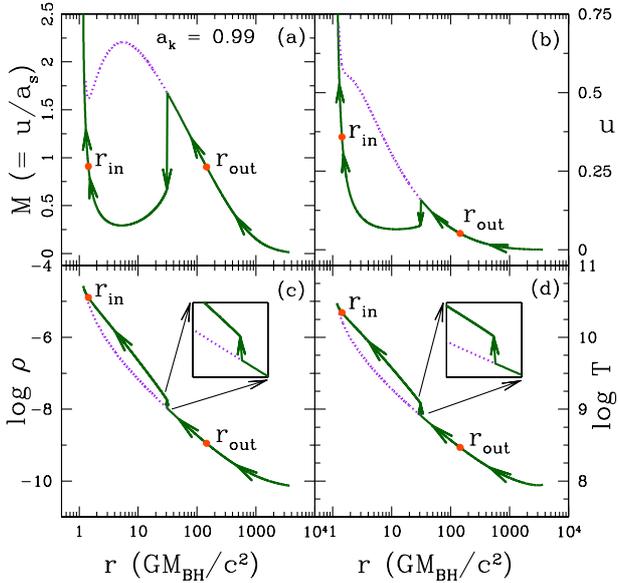}
	\end{center}
	\vskip -0.3cm
	\caption{Typical accretion solution where the variation of (a) Mach number ($M=u/a_s$), (b) velocity (u), (c) density ($\log \rho$), and temperature ($\log T$) are plotted as function of radial distance ($r$). See text for details. 
	}
	\vskip -0.4cm
	\label{fig:r_M}
\end{figure}

In Fig. \ref{fig:r_M}, we present a typical accretion solution containing shock around a rapidly rotating black hole ($a_{\rm k}=0.99$). Here, we fix the global flow parameters, namely viscosity parameter $\alpha = 0.01$ and accretion rate ${\dot m}=0.5$, respectively, and choose the local flow parameters at the inner critical point ($r_{\rm in} = 1.43356$) as ${\cal E}_{\rm in}=1.00741$, angular momentum $\lambda_{\rm in}=1.99$. In the figure, we show the variation of (a) Mach number ($M=u/a_s$), (b) velocity ($u$), (c) density ($\log \rho$), and (d) temperature ($\log T$, in Kelvin) of the flow as function of radial coordinate ($r$), where outer critical point ($r_{\rm out}=144.66855$) and shock location ($r_s=30.80356$) are marked. Because of the shock, inflowing matter undergoes discontinuous transition from supersonic to subsonic branch that yields the jump of density ($\rho$) and temperature ($T$) in the post-shock region ($i.e.$, PSC). Note that after crossing $r_{\rm out}$, flow may eventually enter into the black hole following the dotted curve provided shock conditions are not favorable. In each panel, we indicate the overall direction of the flow motion using arrows. 

In order to explain the observables of ULX sources, we calculate the disc luminosity for a given accretion solution considering gravitational red-shift (${\cal G}$) as $L=4 \pi \int_{r_i}^{r_f} {\cal G} Q^{-}r H dr,$
where $r_i$ refers to the location just outside the horizon ($r_h$),
$r_f$ stands for the outer edge of the disc ($\gtrsim r_{\rm out}$), and $Q^{-}$ denotes the total cooling rates expressed in units of erg cm$^{-3}$ s$^{-1}$. Here, following \cite{Shapiro-Teukolsky1986}, we coarsely approximate ${\cal G}=1-\frac{2r}{(r^2+a_{\rm k}^2)}$ for simplicity.
Further, we examine the QPO features that may originate due to the modulation of the shock front at infall time scales, where infall time is estimated as the time required to accrete the infalling material on to the gravitating object from the shock front. As \cite{Molteni-etal1996} pointed out that the post-shock flow can exhibit non-steady behavior because of the resonance oscillation that happens when the infall timescale is comparable to the cooling time scale of the post-shock flow ($i.e.$, PSC). Since the modulation of PSC in general exhibits Quasi-periodic variations, we estimate the frequency of such modulation as $\nu_{\rm QPO}\sim 1/t_{\rm infall}$, where $t_{\rm infall}=\int_{r_{\rm s}}^{r_i}u_{+}^{-1}dr$, $u_{+}$ is the post-shock velocity \citep{Aktar-etal2015,Dihingia-etal2019}. Employing the above considerations, we calculate the disk luminosity and oscillation frequency for the accretion solution presented in Fig. \ref{fig:r_M} and obtain as $L=1.45 \times 10^{33} \left( \frac{M_{\rm BH}}{M_\odot} \right)^3$ erg s$^{-1}$, and  $\nu_{QPO} = 499.23 \left( \frac{M_{\odot}}{M_{\rm BH}} \right)$ Hz, respectively for ${\dot m} = 0.5$ and $\alpha = 0.01$.

\begin{figure}
	\begin{center}
		\includegraphics[scale=0.4]{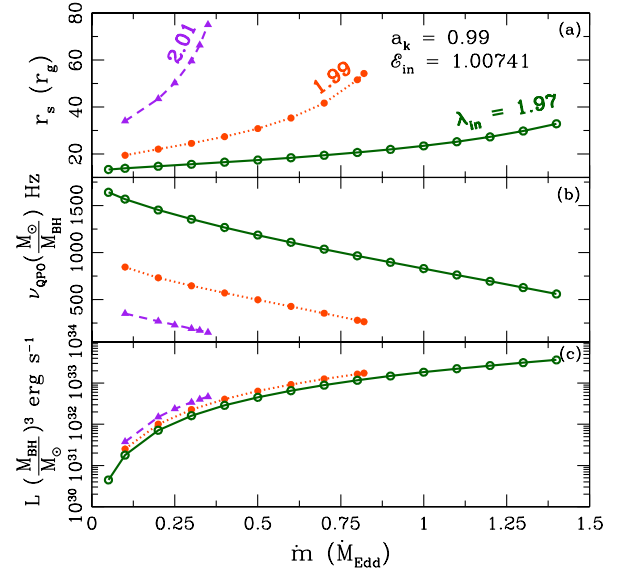}
	\end{center}
	\vskip -0.3 cm
	\caption{ Variation of (a) shock location ($r_{\rm s}$), (b) QPO frequency ($\nu_{\rm QPO}$), and (c) disk luminosity ($L$) as function of ${\dot m}$. See text for details.
	}
	\vskip -0.4 cm
	\label{fig:mdot_xsnuL}
\end{figure}

Next, we examine the role of ${\dot m}$ in determining the shock location ($r_s$), QPO frequency ($\nu_{\rm QPO}$), and disk luminosity ($L$). The obtained results are depicted in Fig. \ref{fig:mdot_xsnuL}, where we fix $a_{\rm k}=0.99$, $\alpha = 0.01$, and ${\cal E}_{\rm in} = 1.00741$. In Fig. \ref{fig:mdot_xsnuL}a, we present the variation of $r_s$ with ${\dot m}$ where solid (green), dotted (orange) and dashed (purple) curves are obtained for $\lambda_{\rm in} =1.97$, $1.99$, and $2.01$, respectively. We observe that shocks are formed for a wide range of ${\dot m}$ and generally they settle down at larger radii for flows with higher $\lambda_{\rm in}$. In Fig. \ref{fig:mdot_xsnuL}b, we present $\nu_{\rm QPO}$ which is computed using the results presented in the upper panel. As shocks are formed further out for higher $\lambda_{\rm in}$, the corresponding $\nu_{\rm QPO}$ are yielded at lower values. In Fig. \ref{fig:mdot_xsnuL}c, we show the variation of $L$ with ${\dot m}$ for the same solutions depicted in the upper panel. We find that for a given $\lambda_{\rm in}$, $L$ strongly depends on ${\dot m}$ whereas the response of $\lambda_{\rm in}$ on $L$ is relatively weak for a fixed ${\dot m}$.

With this, we perceive that the present formalism is capable to cater $\nu_{\rm QPO}$ and $L$ for their wide range of values. Hence, we employ the present model formalism to examine $\nu_{\rm QPO}$ and $L_{\rm bol}$ of a well studied ULX source IC 342 X-1, and attempt to constrain $M_{\rm BH}$, $a_{\rm k}$, and $\dot m$ of the source, respectively.

\section{Astrophysical implication: IC 342 X-1}

\subsection{Observational features}

\begin{figure*}
	\begin{center}
		\hskip -1.0cm
		\includegraphics[scale=0.65,angle=270]{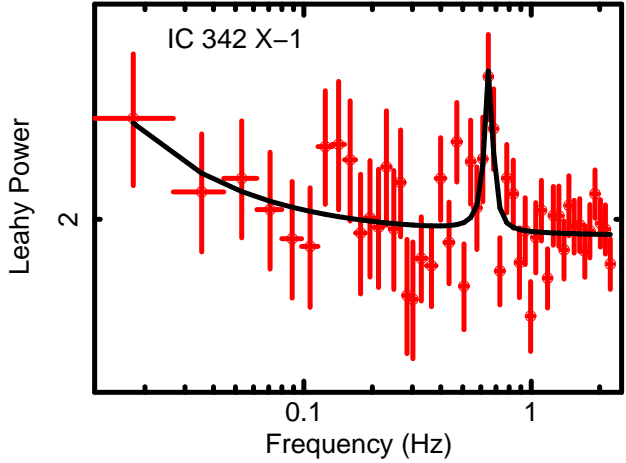}
		\includegraphics[scale=0.55,angle=270]{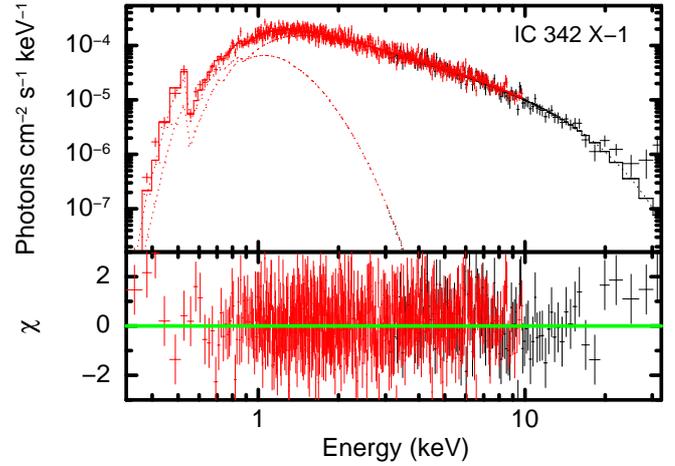}
	\end{center}
	\vskip -0.3cm
	\caption{The PDS (left) of EPIC-pn observation taken during 11 August 2012. The PDS is fitted with a constant, a power-law and a Lorentzian centered at $\sim 645$ mHz. The unfolded energy spectrum (right) of combined fit to the quasi-simultaneous data of {\it NuSTAR-FPMA} and {\it XMM-Netwon/EPIC-pn}. The combined spectrum is fitted with $\texttt {TBabs} \times (\texttt{compTT} + \texttt{diskbb})$ model. See text for details.
			}
	\vskip -0.4 cm
	\label{fig:IC_QPO_Lum}
\end{figure*}

We analyze the quasi-simultaneous observations of IC 342 X-1 carried out on 11 August 2012 with {\it XMM-Newton} and {\it NuSTAR} observatories. We follow the procedures described in \cite{Agrawal-Nandi2015} to generate the lightcurve, spectrum and auxiliary (background, response) files. We use $0.3 -10$ keV {\it XMM-Newton/EPIC-pn} lightcurve with bin size of $0.22$ s to construct the power density spectrum (PDS). We compute the PDS for intervals of $256$ bins and average them over a single frame. We rebinned the final PDS by a geometric factor of $1.04$ in the frequency space. The PDS exhibits a Lorentzian feature at $\sim 645$ mHz. We fit the PDS using a power-law ($ \propto \nu^{-\alpha}$, $\alpha$ being the index), a constant (to represent the Poisson noise) and a Lorentzian (to represent the QPO). Fig. \ref{fig:IC_QPO_Lum} (left) shows the PDS of IC 342 X-1 along with the fitted model. The centroid frequency of QPO ($\nu_{\rm QPO}$) is obtained as $645 \pm 20$ mHz with Q factor $\sim 11$ and significance $\sim 3.8 \sigma$ (see Table 1).

We carry out the spectral analysis using the quasi-simultaneous {\it XMM-Newton} ($0.3 - 10$ keV) and {\it NuSTAR} ($3 - 30$ keV) data. The combined spectrum is fitted with various model combinations available in {\it XSPEC}. We proceed with physically motivated Comptonized model $i.e.$, $\texttt{TBabs} \times (\texttt{compTT} + \texttt{diskbb})$ to extract the spectral parameters. Details of spectral modeling were presented in \citet{Agrawal-Nandi2015}. In Fig. \ref{fig:IC_QPO_Lum} (right), the unfolded energy spectrum is shown along with the residuals (bottom panel). Considering the recent measurement of the source distance $D \sim 3.45$ Mpc \citep{Wu-etal2014} and flux estimation (see Table 1), we calculate the bolometric luminosity ($L_{\rm bol} = 4 \pi D^2 F_{\rm bol}$) as $(7.59 \pm 0.57) \times 10^{39}$ ergs s$^{-1}$, where $F_{\rm bol}$ being the bolometric flux. The model fitted (both temporal and spectral) and computed parameters are summarized in Table 1.

\begin{table}
	\centering
	\caption{Model fitted temporal and spectral parameters for IC 342 X-1. $F_{\rm bol}$ and $L_{\rm bol}$ are computed in $0.1-100$ keV energy range.}
	\begin{tabular}{l|c|c|}
		\hline
		Features & Parameters & Values\\
		\hline
		Timing & $\nu_{\rm QPO}$ (mHz) &  $645 \pm 20$\\ 
		 & $Q ~ (\nu_{\rm QPO}/{\rm FWHM})$ & $11$ \\
		 & $\sigma^\dagger$ & $3.8$ \\\hline
		Spectral & nH ($10^{22}$ atoms/cm$^2$)   & $0.65\pm0.05$ \\
		& $kT_e$ (keV) & $3.3 \pm 0.18$ \\ 
		& $\tau$ & $13.45 \pm 0.65$ \\
		& K ($\times 10^{-4}$)    & $3.51 \pm 0.3$ \\
		& $kT_{in}$ (keV) & $0.23 \pm 0.02$ \\
		& $N_{disk}$ & $23^{+26}_{-11}$ \\
		& $\chi^2$/dof & $638/637$ \\ \hline
		Estimated  & $F_{\rm bol}$ ($\times 10^{-12}$ ergs/s/cm$^2$)  & $5.36 \pm 0.32$ \\
	    & $L_{\rm bol}$ ($\times 10^{39}$ ergs/s) & $7.59 \pm 0.57$ \\
		\hline
	\end{tabular}
	 \justify{$^\dagger$ The QPO significance ($\sigma$) is computed as the ratio of Lorentzian normalization to its negative error \cite[see][and references therein]{Sreehari-etal2019}}.
\end{table}

\subsection{Constraining mass, spin and accretion rate}

\begin{figure}
	\vskip -0. cm
	\begin{center}
	\includegraphics[scale=0.34]{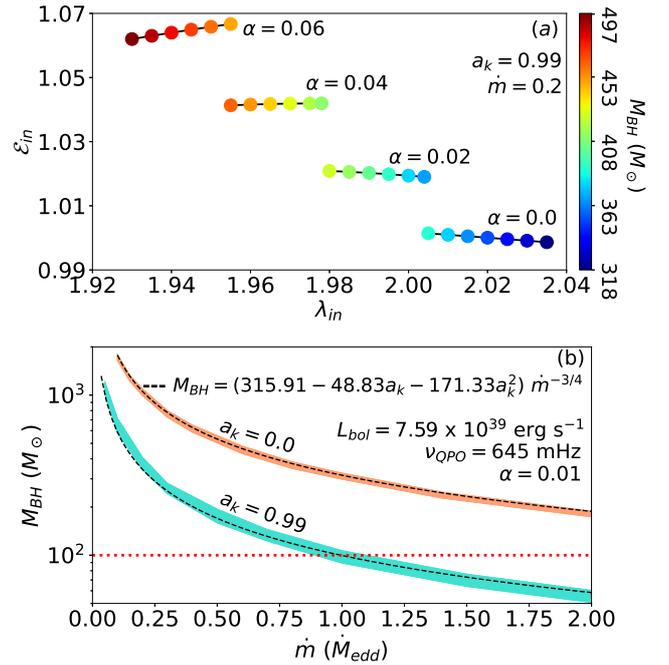}
	\end{center}
\vskip -0.7 cm
	\caption{(a) Variation of ${\cal E}_{\rm in}$ with $\lambda_{\rm in}$ that results observed $L_{\rm bol}$ and $\nu_{\rm QPO}$ for IC 342 X-1 source for different mass range as indicated using colorabar. Here, we choose $a_{\rm k} = 0.99$ and ${\dot m}=0.2$, and $\alpha$ values are marked. (b) Correlation between $\dot m$ and $M_{\rm BH}$ for different $a_{\rm k}$. Regions shaded using orange and cyan color are for $a_{\rm k}=0.0$, and $0.99$, respectively. Dashed curves denote the fitted function as marked in the figure.  See text for details. 
	}
	\vskip -0.4 cm
	\label{fig:mdot_MBH_ak}
\end{figure}

To infer $L_{\rm bol}$ and $\nu_{\rm QPO}$ of IC 342 X-1, we employ the present model formalism. While doing this, we freely vary the flow parameters, namely $\lambda_{\rm in}$ and ${\cal E}_{\rm in}$, to compute the shocked accretion solutions for a given set of parameters ($a_{\rm k}, {\dot m}, \alpha, M_{\rm BH}$), and obtain the solution that yields the observed $L_{\rm bol}$ and $\nu_{QPO}$ for IC 342 X-1 source. The obtained results are depicted in Fig. \ref{fig:mdot_MBH_ak} (a), where for a given viscosity parameter $\alpha$, we show the interplay among $\lambda_{\rm in}$, ${\cal E}_{\rm in}$ and $M_{\rm BH}$ that provides the observed $L_{\rm bol}=7.59 \times 10^{39}$ ergs s$^{-1}$ and $\nu_{\rm QPO} = 645$ mHz of IC 342 X-1 source. Here, we choose $a_{\rm k}=0.99$, and ${\dot m} = 0.2$. In the figure, we mark the $\alpha$ values and indicate the mass ($M_{\rm BH}$) range using the colorbar. It is clear that as $\alpha$ is increased, $\lambda_{\rm in}$ is shifted to the lower values whereas ${\cal E}_{\rm in}$ moved to the higher energy domain.

In Fig. \ref{fig:mdot_MBH_ak} (b), we show the correlation between the source mass ($M_{\rm BH}$) and the accretion rate (${\dot m}$) for $\alpha = 0.01$ that delineate $L_{\rm bol}$ and $\nu_{\rm QPO}$ for IC 342 X-1 source. Here, we compare the results for non-rotating ($a_{\rm k}=0.0$) and rapidly rotating ($a_{\rm k}=0.99$) black hole. For a given ${\dot m}$ and $a_{\rm k}$, we find a range of $\lambda_{\rm in}$, ${\cal E}_{\rm in}$, and $M_{\rm BH}$ that provides the observed $L_{\rm bol}$ and $\nu_{\rm QPO}$ for IC 342 X-1 source. Using these results, we empirically obtain a functional form of $M_{\rm BH}= (315.91-48.83a_{\rm k}-171.33a^2_{\rm k}){\dot m}^{-3/4}$, which is characterized as seemingly exponential with the accretion rate (${\dot m}$) shown by the dashed curves.  We observe that for $a_{\rm k} = 0.0$, IC 342 X-1 seems to accrete matter both at sub- and super-Eddington limits depending on its mass ($174 \lesssim \frac{M_{\rm BH}}{M_{\odot}}\lesssim 1783$). Similarly, when $a_{\rm k}=0.99$, we obtain the corresponding mass range of the source as $55 \lesssim \frac{M_{\rm BH}}{M_{\odot}}\lesssim 1198$ for ${\dot m \lesssim 2}$.

\section{Discussion and Conclusions}

\label{sec:con}

In this letter, we develop and discuss a model of relativistic, viscous, advective accretion flows around ULX sources that harbor black holes as central objects. To describe the effect of gravity, we adopt a newly developed effective potential \citep{Dihingia-etal2018}, that satisfactorily mimics the spacetime geometry around the rotating black hole. We solve the basic equations that govern the flow motion and calculate the global accretion solutions in presence of shocks (Fig. 1). Indeed, the shock induced accretion solutions are presumably promising as they are capable to explain the observational signatures of the black hole X-ray sources \citep[and references theein]{Iyer-etal2015,Sreehari-etal2019}.

We find that the shock induced global accretion solutions exist
for a wide range of flow parameters, namely ${\cal E}_{\rm in}$, $\lambda_{\rm in}$, $\alpha$, and ${\dot m}$, respectively. Considering the resonance oscillations \citep{Molteni-etal1996}, we inductively compute the QPO frequency ($\nu_{\rm QPO}$) associated with the infall time scale of the material parcel within the PSC. For these accretion solutions, we also compute the disc luminosity ($L$) by integrating the rate of energy loss over the entire disc. We observe that depending on the flow parameters, both $\nu_{\rm QPO}$ and $L$ span over their extensive range of values (Fig. 2). 

Further, we examine the observational findings of IC 342 X-1 which exhibits QPO feature with centroid frequency at $645\pm20$ mHz and bolometric luminosity $L_{\rm bol} = 7.59 \times 10^{39}$ erg s$^{-1}$ (see Fig. 3, Table 1). Employing the present model formalism, we compute $\nu_{\rm QPO}$ and $L_{\rm bol}$ for IC 342 X-1 by tuning the input parameters ($i.e.$, ${\cal E}_{\rm in}$, $\lambda_{\rm in}$, $\alpha$, ${\dot m}$ and $a_{\rm k}$). We observe that matter can accrete onto IC 342 X-1 both at sub- and supper-Eddington accretion rates depending on the source mass irrespective of spin parameter ($a_{\rm k}$). Considering these findings, we infer the possible mass range of the IC 342 X-1 based on its spin and accretion rate, and find that for $a_{\rm k}=0.99$ and ${\dot m} \lesssim 2$, $55 \lesssim \frac{M_{\rm BH}}{M_\odot} \lesssim 1198$ and for $a_{\rm k}=0.0$ and ${\dot m} \lesssim 2$, $174 \lesssim \frac{M_{\rm BH}}{M_\odot} \lesssim 1783$. In addition, using the above insights, we empirically find that $M_{\rm BH} \sim {\dot m}^{-3/4}$ for a given black hole spin $a_{\rm k}$ (see Fig. 4b).

Finally, we indicate that the present formalism satisfactorily complies the findings of \cite{Agrawal-Nandi2015,Shidatsu-etal2017}, where they reported the presence of a massive stellar mass black hole ($M_{\rm BH} < 65 M_\odot$) in IC 342 X-1.
Considering this, we ascertain that the central source in IC 342 X-1 seems to be rapidly rotating and it perhaps accretes matter at super-Eddington accretion rate. 

We mention the limitation of our work that the accretion solutions largely remain optically thin except in the vicinity of the black hole horizon for high accretion rate (${\dot m} \sim 2$) and therefore, we argue that the basic conclusion of the present paper will remain qualitatively unaltered.

\section*{Data availability statement}

The data underlying this article are available in the published literature as well as at High Energy Astrophysics Science Archive Research Centre (HEASARC) facility, located at NASA-Goddard Space Flight Centre.

\section*{Acknowledgments}
We thank the anonymous referee for valuable comments and suggestions that helped to improve the manuscript. SD thanks Science and Engineering Research Board (SERB), India for support under grant MTR/2020/000331. AN and VKA thanks GH, SAG; DD, PDMSA and Director, URSC for encouragement and continuous support to carry out this research. IKD thanks the financial support from Max Planck partner group award at Indian Institute Technology of Indore (MPG-01).



\label{lastpage}

\end{document}